\documentclass[conference]{IEEEtran}\IEEEoverridecommandlockouts
\usepackage{etoolbox}
\usepackage{algorithmic}
\usepackage{algorithm}

\makeatletter
\patchcmd{\@makecaption}
  {\scshape}
  {}
  {}
  {}
\makeatletter
\patchcmd{\@makecaption}
  {\\}
  {.\ }
  {}
  {}
\makeatother
 \usepackage{amsmath,amssymb}
 \usepackage{subfigure}
 \usepackage{graphicx,graphics,color,psfrag}
 \usepackage{cite,balance}
 \usepackage{caption}
 \allowdisplaybreaks
 \usepackage{algorithm}
 \usepackage{accents}
 \usepackage{amsthm}
 \usepackage{bm}
 \usepackage{algorithmic}
 \usepackage[english]{babel}
 \usepackage{multirow}
 \usepackage{enumerate}
 \usepackage{cases}
 \usepackage{stfloats}
 \usepackage{dsfont}
 \usepackage{color,soul}
 \usepackage{amsfonts}
 \usepackage{cite,graphicx,amsmath,amssymb}
 \usepackage{subfigure}
 \usepackage{fancyhdr}
 \usepackage{hhline}
 \usepackage{graphicx,graphics}
 \usepackage{array,color}
 \usepackage{amsmath}
\usepackage{float}
\usepackage{amssymb}
\usepackage{amsmath}
\usepackage{amsthm}
\usepackage{amsfonts}
\usepackage{graphicx}

\usepackage{epstopdf}
\usepackage{cite}
\usepackage{amsmath,bm}
\usepackage{subfigure}
\usepackage{graphicx}
\usepackage{color}
\usepackage{graphicx}
\usepackage{calc}
\usepackage{caption}

\newtheorem{remark}{\textbf{Remark}}

\begin{document}

\title{Fusion-Based Multi-User Semantic Communications for Wireless Image Transmission over Degraded Broadcast Channels}
\author{Tong Wu, Zhiyong Chen, Meixia Tao, Bin Xia, Wenjun Zhang\\
		Cooperative Medianet Innovation Center, Shanghai Jiao Tong University, Shanghai, China\\
		Email: \{wu\_tong, zhiyongchen, mxtao, bxia, zhangwenjun\}@sjtu.edu.cn}

\maketitle
\begin{abstract}
Degraded broadcast channels (DBC) are a typical multi-user communications scenario. There exist classic transmission methods, such as superposition coding with successive interference cancellation, to achieve the DBC capacity region. However, semantic communications method over DBC remains lack of in-depth research. To address this, we design a fusion-based multi-user semantic communications system for wireless image transmission over DBC in this paper. The proposed architecture supports a transmitter extracting semantic features for two users separately, and learns to dynamically fuse these semantic features into a joint latent representation for broadcasting. The key here is to design a flexible image semantic fusion (FISF) module to fuse the semantic features of two users, and to use a multi-layer perceptron (MLP) based neural network to adjust the weights of different user semantic features for flexible adaptability to different users channels. Experiments present the semantic performance region based on the peak signal-to-noise ratio (PSNR) of both users, and show that the proposed system dominates the traditional methods.

\end{abstract}
\section{Introduction}\label{I}
In recent years, semantic communications have received significant attention from both industry and academia. With the help of artificial intelligence (AI), semantic communications can extract the semantic information from the original data, and further transmit it, thereby significantly improving communication efficiency \cite{gundu2019}. Therefore, semantic communications have been considered a promising solution for the sixth-generation (6G) wireless networks \cite{shen2}.

Several studies have been conducted on semantic communications for different types of original information, such as text \cite{text2021, text2022}, image \cite{image2022, image2022tao}, and video \cite{gundu2022video, video2022, video2022-2}. For text transmission, a deep learning-based semantic communication system is proposed in \cite{text2021}, named DeepSC, which has an advantage in the low signal-to-noise ratio (SNR) regime. For image transmission, a deep learning-based semantic image coding method is designed in \cite{image2022} to encode images beyond pixel level. For video transmission, the end-to-end joint source-channel coding (JSCC) video transmission scheme is first proposed in \cite{gundu2022video}. Then, \cite{video2022} designs a novel deep joint source-channel coding approach to achieve wireless video transmission, which can outperform traditional wireless video coded transmission schemes.

It is worth noting that previous works mainly focus on point-to-point semantic communications, while research on multiuser semantic communications is relatively limited. In \cite{noma}, a heterogeneous semantic and bit communication framework is designed for multiple access channels that utilizes a method called semi-nonorthogonal multiple access (NOMA) and achieves better performance than the classic NOMA system. Meanwhile, a novel joint image compression and transmission scheme for the multi-user uplink scenario is presented in \cite{gundu2022MAC}, which utilizes NOMA and incorporates deep neural networks (DNNs) into the transmitters. For broadcasting channels, a one-to-many scheme is proposed in \cite{Hu} for text transmission, where the transmitter concatenates these texts together and extracts their semantic features for transmission. For relay channels, a semantic-and-forward scheme is first designed in \cite{relay} to address the heterogeneous background knowledge problem. Then, a novel deep joint source-channel coding scheme for image transmission over a half-duplex cooperative relay channel is presented in \cite{gundu2022relay}.

Actually, multi-user semantic communications are not simply point-to-point semantic communications but require corresponding design for multi-user channels. Motivated by this, we consider a degraded broadcast channel (DBC) in this paper, which is a typical multiuser communication scenario. There is a transmitter and multiple users located in different geographical locations. The capacity region of DBC is well-known, and there are many traditional transmission methods, such as superposition coding with successive interference cancellation, time division (TD) and frequency division (FD), to achieve the DBC capacity region. However, in semantic communications, can the semantic information of two users be deeply integrated through AI networks rather than simply transmitted together using superposition coding?


To address this issue, we propose a fusion-based multi-user semantic communications system for wireless image transmission over two-user DBC. In the proposed architecture, a transmitter can extract and fuse the semantic features as a joint latent representation of both users. The worse user can only decode its own image from the joint representation, while the better user reconstructs the other image first and then obtains its own image based on the reconstructed image. To deal with the fuse of semantic features for two users, we design a flexible image semantic fusion (FISF) scheme to dynamically control the weight of two users' semantic features in the joint latent representation by using a neural network based on multi-layer perceptron (MLP). Meanwhile, to adapt the respective semantic features to the respective channels with different SNRs, the proposed FISF scheme uses the attention mechanism with channel state information (CSI) to adapt the different channel condition \cite{chenwei,wuhaotian}. Numerical results based on real-world datasets show that the proposed system can significantly improve the peak signal-to-noise ratio (PSNR) of the images for both users.

\section{System Model and Strategy Design}
In this section, we propose a semantic communication strategy for wireless image transmission over the degraded broadcast channel. The system consists of a transmitter and two users, where two image messages $\mathbf{s}_1$ and $\mathbf{s}_2$ are expected to be delivered to the two users through semantic communication.

\begin{figure*}[t]
  \begin{center}
    \includegraphics[width=16cm,scale=2.00]{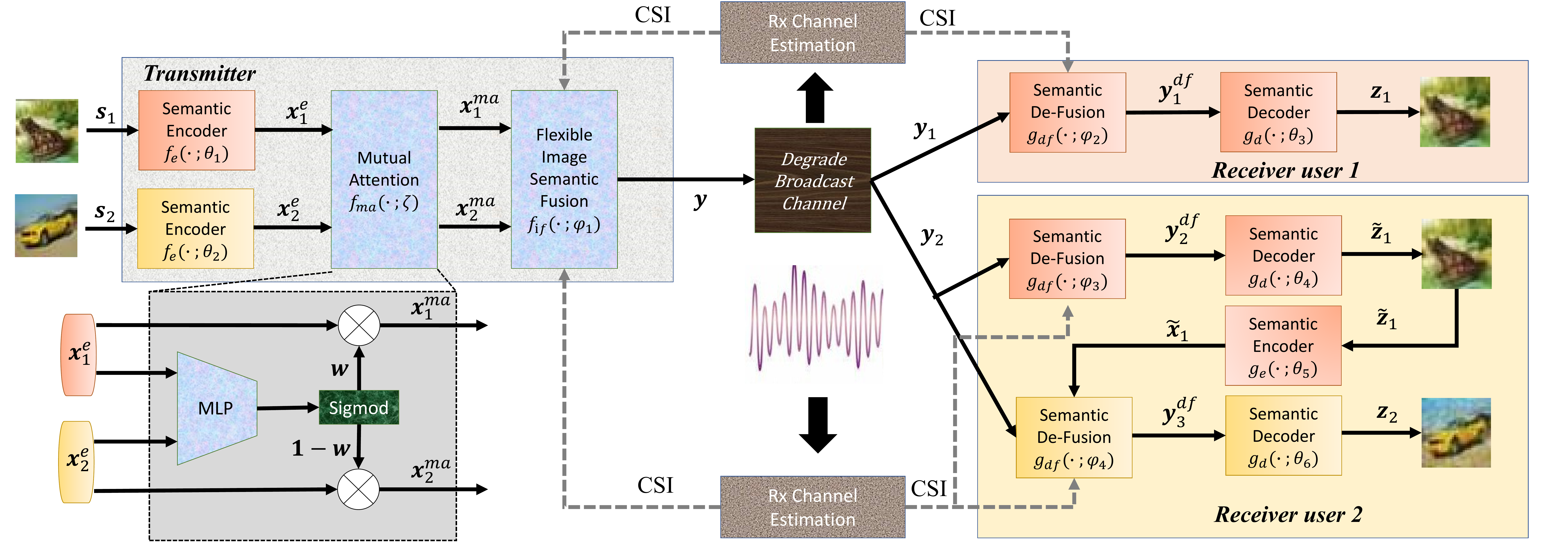}
  \end{center}
  \caption{\small{The structure of the proposed degraded broadcast semantic communication system}}
  \label{model}
\end{figure*}
\subsection{System Overview}
As shown in Fig. \ref{model}, two semantic encoders (SE) $f_{e}(\cdot;{\theta_1})$ and $f_{e}(\cdot;{\theta_2})$ can extract the image features $\mathbf{x}_{1}^{e}$ and $\mathbf{x}_{2}^{e}\in{\mathbb{R}^{l}}$ from the two source images $\mathbf{s}_1$ and $\mathbf{s}_2\in{\mathbb{R}^{h\times n\times 3}}$ respectively, where $l$ is the output dimension of the semantic encoders, $h$ and $n$ denote the height and width of the image, and 3 is the color channels $R$, $G$ and $B$. $f_{e}$ is the module strcuture and $\theta_i$, $i=1,2$, is learning parameter. $\mathbf{x}_{1}^{e}$ and $\mathbf{x}_{2}^{e}$ are then fed into a mutual attention (MA) module $f_{ma}(\cdot;\zeta  )$, which first computes a element-wise weight $\mathbf{w}\in\mathbb{R}^{l}$. The relationship between $\mathbf{x}_1^{e}$ and $\mathbf{x}_2^{e}$ is associated with $\mathbf{w}$. Then, the input features can be fused with $\mathbf{w}$ in the element-wise product as
\begin{gather}\label{mutual1}
  \mathbf{x}_1^{ma}=\mathbf{x}_{1}^{e}\odot \mathbf{w},\\
  \mathbf{x}_{2}^{ma}=\mathbf{x}_{2}^{e}\odot \mathbf{(1-w)},
\end{gather}
where $\mathbf{x}_{1}^{ma}$ and $\mathbf{x}_{2}^{ma}$ are the outputs of the $f_{ma}(\cdot;\zeta  )$. We then design FISF module $f_{if}(\cdot;{\varphi _1})$ to fuse $\mathbf{x}_{1}^{ma}, \mathbf{x}_2^{ma}\in\mathbb{R}^{l}$ into a joint latent representation $\mathbf{y}\in\mathbb{R}^k$ by using a fusion ratio $\alpha$ for controlling the reconstruction quality of two users, where $k$ is the number of channel uses. The transmitter also inform FISF with the channel state information and thus the output $\mathbf{y}$ can fit the degrade broadcast channel. 

We consider two distant users, one with Gaussian noise power $\sigma_1^2$ and the other with Gaussian noise power $\sigma_2^2$. Without loss of generality, we assume that $\sigma_1^2>\sigma_2^2$. The received signals of the two users are $\mathbf{y}_1=\mathbf{y}+\mathbf{n}_1$ and $\mathbf{y}_2=\mathbf{y}+\mathbf{n}_2$ respectively, where $\mathbf{n}_1$ and $\mathbf{n}_2$ are Gaussian noise with noise power $\sigma_1^2$ and $\sigma_2^2$, respectively. We call the user with noise power $\sigma_1^2$ as the worse user and the other as the better user. Similar to the traditional DBC, the worse user can only decode its own message. Therefore, at the worse user, $\mathbf{y}_1$ is fed into the de-fusion(DF) module $g_{df}(\cdot;{\varphi_2})$ with $\alpha$ and CSI, yielding the output $\mathbf{y}_1^{df}\in\mathbb{R}^k$. The user then performs a semantic decoder (SD) $g_{d}(\cdot;{\theta _2})$ to reconstruct the image $\mathbf{s}_1$ as $\mathbf{z}_1\in\mathbb{R}^{h\times n\times 3}$.

The better user first reconstructs the image $\mathbf{s}_1$ and then reconstructs its own image $\mathbf{s}_2$ based on $\mathbf{s}_1$. Specifically, upon receiving $\mathbf{y}_2$, the user can reconstruct the image $\mathbf{s}_1$ as $\tilde{\mathbf{z}}_1$ by performing the DF module $g_{df}(\cdot;{\varphi_3})$, the SD module $g_{d}(\cdot;\theta_4)$ and the SE module $g_e(\cdot;\theta_5)$. Therefore, the user can obtain the features of $\tilde{\mathbf{z}}_1$ as $\tilde{\mathbf{x}}_1\in\mathbb{R}^{l}$. Meanwhile, with $g_{df}(\cdot;{\varphi_4})$ and $\tilde{\mathbf{x}}_1$, the user can reconstruct the image $\mathbf{s}_2$ as  $\mathbf{z}_2\in\mathbb{R}^{h\times n\times 3}$ by performing the SD module $g_d(\cdot;\theta_6)$.

\begin{figure*}[t]
  \begin{center}
    \includegraphics*[width=14cm]{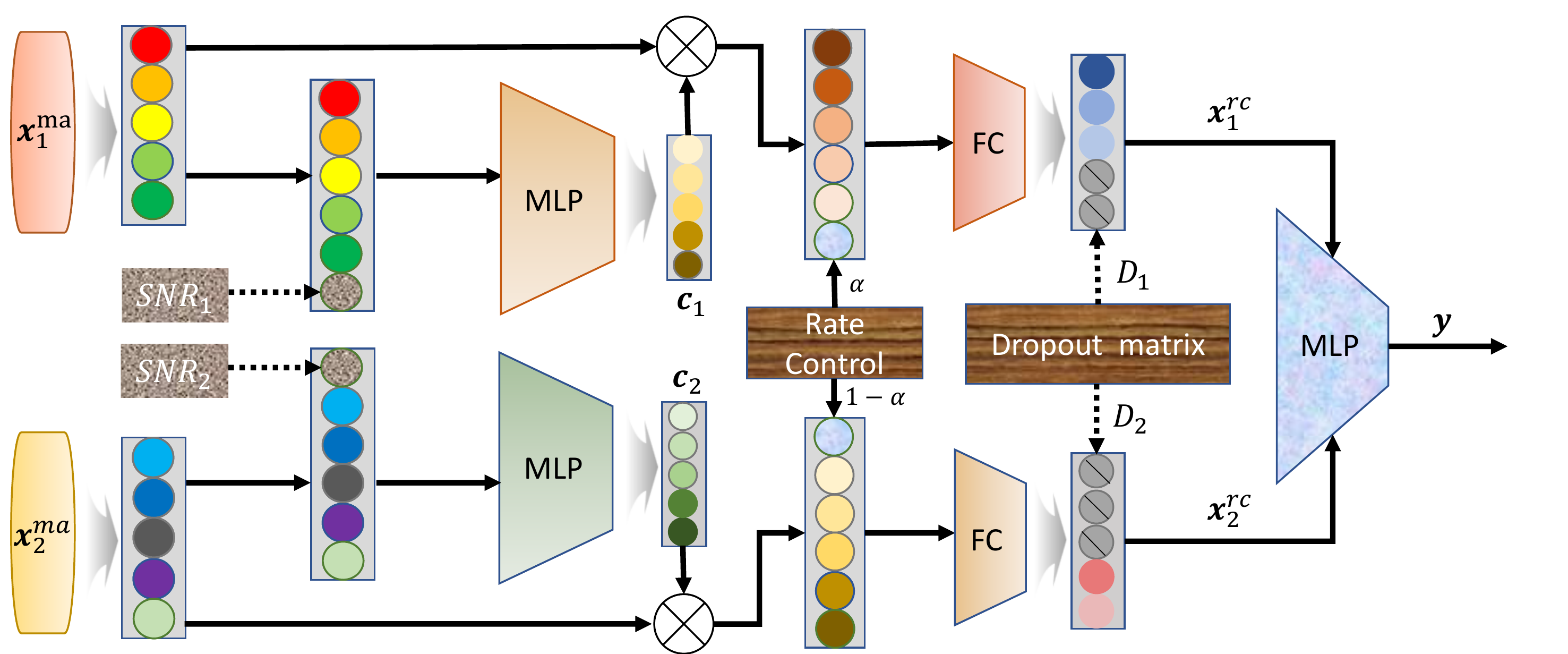}
  \end{center}
    \caption{\small{The structure of flexible image semantic fusion}}
    \label{DBCA}
\end{figure*}
We note here that neural networks are utilized for the SE module, the MA module and the SD in this paper. In the following, we detail the design of the FISF module.

\subsection{Flexible Image Semantic Fusion Module}
For DBC, the channel input $\mathbf{y}$ is a joint latent represenatation of $\mathbf{s}_1$ and $\mathbf{s}_2$. The component of $\mathbf{y}$ from $\mathbf{s}_1$ is required to fit the worse channel because the worse user only needs to reconstruct $\mathbf{s}_1$ from the received signal $\mathbf{y}_1$. Meanwhile, all the componets of $\mathbf{y}$ is required to fit the better channel because the better user requires to reconstruct $\mathbf{s}_1$ and $\mathbf{s}_2$ from $\mathbf{y}_2$. Therefore, how to flexibly fuse $\mathbf{s}_1$ and $\mathbf{s}_2$ to $\mathbf{y}$ for adapting to both channel states is crucial to the system design.

In this paper, we develop a flexible image semantic fusion strategy to fuse $\mathbf{s}_1$ and $\mathbf{s}_2$ to flexibly adapt the different channel conditions and dynamic control the weight of $\mathbf{s}_1$ and $\mathbf{s}_2$ in $\mathbf{y}$. The architecture of the proposed FISF module is shown in Fig. \ref{DBCA}. First, the output $\mathbf{x}_1^{ma}$ of MA is combined with the SNR of the worse channel and fed into an MLP to generate a vector $\mathbf{c}_1\in\mathbb{R}^{l}$. Likewise, we can obtain a vector $\mathbf{c}_2\in\mathbb{R}^{l}$ from the output $\mathbf{x}_2^{ma}$ of MA for the better channel. As a result, $\mathbf{c}_1$ and $\mathbf{c}_2$ are the attention masks that contain the image feature and the channel station information. We can adjust the image feature by scaling the attention mask to transmit the image feature in a more robust form in the channel as following
\begin{gather}
  \mathbf{x}_1^{ca}=\mathbf{x}_1^{ma}  \odot \mathbf{c}_1,\ 
  \mathbf{x}_2^{ca}=\mathbf{x}_2^{ma}\odot \mathbf{c}_2.
\end{gather}

Next, to dynamic control the weight of $\mathbf{s}_1$ and $\mathbf{s}_2$ in $\mathbf{y}$,  $\mathbf{x}_1^{ca}$ is combined with a fusion ratio $\alpha$ and then fed into full-connected (FC) layers. The FC can sort the semantic vector and put the important parts in the front of the vector. The output of the FC is multiplied by a non-square identity matrix $\mathbf{D}_1=[I_{\lfloor 2\alpha k\rfloor},O]\in\mathbb{R}^{\lfloor \alpha l\rfloor \times l}$ to produce $\mathbf{x}_1^{rc}\in\mathbb{R}^{\alpha l}$, where $\lfloor\cdot \rfloor$ is round-down function. Here, $I_j$ is j-dimension identity matrix and $O$ is zero matrix. Similarly, we have $\mathbf{x}_2^{rc}\in\mathbb{R}^{(1-\alpha) l}$ based on $\mathbf{D}_2=[I_{\lceil 2(1-\alpha) k\rceil},O]\in\mathbb{R}^{\lceil (1-\alpha) l\rceil \times l}$, where  $\lceil\cdot \rceil $ is round-up function. This process is described as
 \begin{gather}
  \mathbf{x}_1^{rc}=\mathbf{D}_1*(\mathbf{A}_1[(\mathbf{x}_1^{ca})^T,\alpha]+\mathbf{b}_1),\\
  \mathbf{x}_2^{rc}=\mathbf{D}_2*(\mathbf{A}_2[(\mathbf{x}_2^{ca})^T,1-\alpha]+\mathbf{b}_2),
 \end{gather}
 where $\mathbf{A}_i$ and $\mathbf{b}_i$ are affine function parameter and their bias of FCs, respectively. Then, $\mathbf{x}_1^{rc}$ and $\mathbf{x}_2^{rc}$ are passed through a MLP to generate $\mathbf{y}\in\mathbb{R}^k$. According to the definition of MLP, we have
 \begin{equation}
  \mathbf{y}=\tanh(\mathbf{Q}*[(\mathbf{x}_1^{rc})^T,(\mathbf{x}_2^{rc})^T]^T+\mathbf{d}),\label{mlp_y}
  \end{equation}
where $\mathbf{Q}\in\mathbb{R}^{k\times l}$ and $\mathbf{d}\in\mathbb{R}^{k}$ are learning parameters. Here, we use tanh activation as the activation function for this MLP. Based on (\ref{mlp_y}), the $i$-th transmitted symbol can be expressed as 
 \begin{gather}
  y_i=\tanh(\sum_{j=1}^{\lfloor \alpha l \rfloor} q_{ij}{x_{1,j}^{rc}}+\sum_{j=\lfloor \alpha l \rfloor+1}^{l} q_{ij}{x_{2,(j- \lfloor \alpha l\rfloor)}^{rc}}+d_i).\label{mlp_y_t}
  \end{gather}
Finally, we can perform the power normalization for $\mathbf{y}$ and deliver it over the channel.

\begin{remark}
Equ. (\ref{mlp_y_t}) reveals that we can dynamic control the weight of $\mathbf{s}_1$ (or $\mathbf{s}_2$) in $\mathbf{y}$ by adjusting $\alpha$ to obtain different decoding performance of two users. We can also see from (\ref{mlp_y_t}) that it is different with the superposition coding scheme in DBC. For the superposition coding, $y_i$ should be $q_ix_{1,i}^{rc}+w_ix_{2,i}^{rc}$, which means $x_{1,i}^{rc}$ and $x_{2,i}^{rc}$ only transmit one time over the channel. However, it is shown in (\ref{mlp_y_t}) that $x_{1,i}^{rc}$ and $x_{2,i}^{rc}$ can be transmitted over the channel multiple times.
\end{remark}

\section{LOSS FUNCTION AND TRAINING METHOD}
We can observe from Fig.\ref{model} that the performance of $\mathbf{z}_1$ and $\mathbf{z}_2$ is dependent on the semantic encoder/decoder and the FISF. SE $f_e(\cdot;\theta)$ and SD $g_d(\cdot;\theta)$ have been well-researched in the point-to-point semantic communication, thus we mainly focus on the loss function design of FISF module in this paper. For the point-to-point system, $\mathbf{s}_{1}$ and $\mathbf{s}_{2}$ can be encoded as $\mathbf{x}^{e2e}_{1}$ and $\mathbf{x}^{e2e}_{2}$, and can be decoded as $\mathbf{z}_{1}^{e2e}$ and $\mathbf{z}_{2}^{e2e}$, respectively.

The worse user only needs to reconstruct $\mathbf{s}_1$. Thus, in this paper, we can design the training object to maximize the conditional mutual information between $\mathbf{z}_1$ and $\mathbf{z}_1^{e2e}$ given $\mathbf{x}^{e2e}_{1}$ and $\mathbf{x}^{e}_1$, as given by
\begin{align}\label{mutual_inforamtion_user1}
  \max\ I(\mathbf{z}_1;\mathbf{z}_{1}^{e2e}|\mathbf{x}^{e}_1,\mathbf{x}^{e2e}_{1}).
\end{align}
It indicates the proposed system tries to output a similar image at the worse user as the excellent image in the point-to-point system. Based on the work of \cite{Su}, this optimization object is hard to achieve. We can achieve a relaxation object by predicting $\hat{\mathbf{z}}_{1}^{e2e}=h_\kappa (\mathbf{z}_1,\mathbf{x}^{e}_1,\mathbf{x}^{e2e}_{1})$ first and then estimating the posterior distribution $\mathcal{P} _{\kappa }(\mathbf{z}_{1}^{e2e}|\hat{\mathbf{z}}_{1}^{e2e})$. The relaxation form can be written as
\begin{align}\label{lossform}
  I(\mathbf{z}_1;\mathbf{z}_{1}^{e2e}|\mathbf{x}^{e}_1,\mathbf{x}^{e2e}_{1})=\sup_{h_\kappa}\  \mathbb{E}_{p(\mathbf{x}_{1}^{e2e})}\ [H({p(\mathbf{z}_{1}^{e2e}|\mathbf{x}_{1}^{e2e})})]\nonumber\\
  +\mathbb{E}_{p(\mathbf{s}_1,\mathbf{x}^{e}_1,\mathbf{x}^{e2e}_{1})}\ [\log\mathcal{P} _{\kappa }(\mathbf{z}_{1}^{e2e}|\hat{\mathbf{z}}_{1}^{e2e})],
\end{align}
where $H(x)$ denotes the entropy of random variable $x$. The first term is regularization to avoid collapse in the point-to-point system and the second term is log-likelihood prediction term for target representation. Because the end-to-end system is well-train, the first term is irrelevant to the training process.

 Therefore, (\ref{mutual_inforamtion_user1}) can be solved by deriving the training loss as
\begin{align}\label{L1}
  \min\ L_1(\theta,\zeta ,\varphi)= -\log\mathcal{P} _{\kappa }(\mathbf{z}_{1}^{e2e}|\hat{\mathbf{z}}_{1}^{e2e})
\end{align}
where $\theta$, $\zeta$, $\varphi$ are all the parameters of the whole network. If the estimating posterior distribution $\mathcal{P}$ is Gaussian distribution, it becomes mean squared error (MSE) loss. If it is Boltzmann distribution, it becomes softmax cross-entropy loss. 

\addtolength{\topmargin}{0.015in}
The better user reconstructs $\mathbf{s}_1$ first and then reconstructs $\mathbf{s}_2$. Likewise, the training object can writen as
\begin{align}\label{mutual inforamtion}
  &\max\ I(\tilde{\mathbf{z}}_1,\mathbf{z}_2;\tilde{\mathbf{z}}_{1}^{e2e},\mathbf{z}_{2}^{e2e}|\mathbf{x}^{e}_1,\mathbf{x}^{e}_2,\mathbf{x}^{e2e}_{1},\mathbf{x}^{e2e}_{2})\nonumber\\
  &=\sup_{h_\kappa  }\  \mathbb{E}_{p(\mathbf{x}_{1}^{e2e},\mathbf{x}_{2}^{e2e})}\ [H({p(\tilde{\mathbf{z}}_{1}^{e2e},\mathbf{z}_{2}^{e2e}|\mathbf{x}_{1}^{e2e},\mathbf{x}_{2}^{e2e})})]\nonumber\\
  &+\mathbb{E}_{p(\mathbf{s}_1,\mathbf{x}^{e}_1,\mathbf{x}^{e2e}_{1},\mathbf{s}_2,\mathbf{x}^{e}_2,\mathbf{x}^{e2e}_{2})}\ [\log\mathcal{P} _{\kappa }(\tilde{\mathbf{z}}_{1}^{e2e},\mathbf{z}_{2}^{e2e}|\hat{\tilde{\mathbf{z}}}_{1}^{e2e},\hat{\mathbf{z}}_{2}^{e2e})].
\end{align}
where $\hat{\mathbf{z}}_{2}^{e2e}=h_\kappa (\mathbf{z}_2,\mathbf{x}^{e}_2,\mathbf{x}^{e2e}_{2})$. Similar to (\ref{lossform}), the secord term is critial. Furthermore, we can prove it as
\begin{align}
  &\mathbb{E}_{p(\mathbf{s}_1,\mathbf{x}^{e}_1,\mathbf{x}^{e2e}_{1},\mathbf{s}_2,\mathbf{x}^{e}_2,\mathbf{x}^{e2e}_{2})}\ [\log\mathcal{P} _{\kappa }(\tilde{\mathbf{z}}_{1}^{e2e},\mathbf{z}_{2}^{e2e}|\hat{\tilde{\mathbf{z}}}_{1}^{e2e},\hat{\mathbf{z}}_{2}^{e2e})]\nonumber
  \\&=\mathbb{E}_{p(\mathbf{s}_1,\mathbf{x}^{e}_1,\mathbf{x}^{e2e}_{1},\mathbf{s}_2,\mathbf{x}^{e}_2,\mathbf{x}^{e2e}_{2})}\ [\log\mathcal{P} _{\kappa }(\tilde{\mathbf{z}}_{1}^{e2e}|\hat{\tilde{\mathbf{z}}}_{1}^{e2e},\hat{\mathbf{z}}_{2}^{e2e})\nonumber\\
  &+\log\mathcal{P} _{\kappa }(\mathbf{z}_{2}^{e2e}|\tilde{\mathbf{z}}_{1}^{e2e},\hat{\tilde{\mathbf{z}}}_{1}^{e2e},\hat{\mathbf{z}}_{2}^{e2e})]\nonumber\\
  &=\mathbb{E}_{p(\mathbf{s}_1,\mathbf{x}^{e}_1,\mathbf{x}^{e2r}_{1},\mathbf{s}_2,\mathbf{x}^{e}_2,\mathbf{x}^{e2e}_{2})}\ [\log\mathcal{P}_{\kappa}(\tilde{\mathbf{z}}_{1}^{e2e}|\hat{\tilde{\mathbf{z}}}_{1}^{e2e})\\
  &+\log\frac{\mathcal{P}_{\kappa}(\hat{\mathbf{z}}_{2}^{e2e}|\hat{\tilde{\mathbf{z}}}_{1}^{e2e},\tilde{\mathbf{z}}_{1}^{e2e})}{\mathcal{P}_{\kappa}(\hat{\mathbf{z}}_{2}^{e2e}|\hat{\tilde{\mathbf{z}}}_{1}^{e2e})}+\log\mathcal{P} _{\kappa }(\mathbf{z}_{2}^{e2e}|\tilde{\mathbf{z}}_{1}^{e2e},\hat{\tilde{\mathbf{z}}}_{1}^{e2e},\hat{\mathbf{z}}_{2}^{e2e})]\nonumber.
\end{align}
The first term indicates that $\mathbf{s}_1$ is required to be reconstructed first without any information about $\mathbf{s}_2$ and the third term shows $\mathbf{s}_2$ should be reconstructed under the condition that $\mathbf{s}_1$ has been reconstructed. The structure of the designed user corresponds to formulation that we first reconstruct $\mathbf{s}_1$ as $\tilde{\mathbf{z}}_1$ and then based on $\tilde{\mathbf{z}}_1$, $\mathbf{z}_2$ is reconstructed. The second term indicates that $\mathbf{s}_1$ has effects on $\mathbf{s}_2$. We design the mutual attention module to address the effects.
Therefore, when designing the loss function, the second term is omitted and the third term only contains $\hat{\tilde{z}}_2^{e2e}$. $\beta $ is given to balance the importance of the two terms. The training loss is designed as
\begin{align}
\min\ L_2(\theta,\zeta ,\varphi)&=-\log\mathcal{P} _{\kappa }(\mathbf{z}_{2}^{e2e}|\hat{\mathbf{z}}_{2}^{e2e})\nonumber\\
&-\beta \log\mathcal{P}_{\kappa}(\tilde{\mathbf{z}}_{1}^{e2e}|\hat{\tilde{\mathbf{z}}}_{1}^{e2e}).
\end{align}

Finally, the whole goal of the proposed system is to minimize $L_1$ and $L_2$ at the same time. It also becomes a multi-criterion problem that aims to find the Pareto optimal points. Scalarization is a standard technique for finding Pareto optimal points. The final problem can be solved by deriving the training loss as
\begin{align}
  &\min\ L(\theta,\zeta ,\varphi)=L_1+\lambda L_2=-\log\mathcal{P} _{\kappa }(\mathbf{z}_{1}^{e2e}|\hat{\mathbf{z}}_{1}^{e2e})\nonumber\\
  &-\lambda \log\mathcal{P} _{\kappa }(\mathbf{z}_{2}^{e2e}|\hat{\mathbf{z}}_{2}^{e2e})-\lambda\beta \log\mathcal{P}_{\kappa}(\tilde{\mathbf{z}}_{1}^{e2e}|\hat{\tilde{\mathbf{z}}}_{1}^{e2e}),
\end{align}
where $\lambda$ is the scalarization parameter. In this paper, we consider the posterior distribution $\mathcal{P}_\kappa$ is Gaussian distribution and therefore the loss $L$ can be computed as
\begin{align}
  L=-||\mathbf{z}_1-\mathbf{s}_1||_2-\lambda||\mathbf{z}_2-\mathbf{s}_2||_2-\lambda \beta||\tilde{\mathbf{z}}_1-\mathbf{s}_1||_2.
\end{align}

When training the model, $\mathbf{s}_1$ and $\mathbf{s}_2$ should be generated from the same dataset $S$ individually. A copy of the dataset $\tilde{S}$ is loaded with shuffling. $\mathbf{s}_1$ comes from the batch of $S$ and $\mathbf{s}_2$ comes from the batch $\tilde{S}$. For the FISF module, the SNR of the worse channel ($SNR_1$) is randomly set in a given range and we then randomly set $\gamma>0$ so that the SNR of better channel ($SNR_2$) is $SNR_1+\gamma$. Fusion rate $\alpha$ is also randomly selected in the range between 0 and 1 with step 0.1. When it is 0 or 1, which means only one source is expected to be delivered, the model degrades to an point-to-point model. The whole system takes $L$ as the loss function and all the parameters are updated jointly according to the loss $L$. The whole training procedures are described in Algorithm \ref{train}.
\begin{algorithm}[t]
  \hspace*{0.02in} {\bf \small{Input:}} 
	\small{Training set $S$, hyper-parameter $\lambda$ and $\beta$.} \\
	\hspace*{0.02in} {\bf \small{Output:}} 
	\small{The trained model with one transmitter and two users.}
	\caption{Training algorithm}
	\label{train}
	\begin{algorithmic}[1]
    \STATE Copy a dataset as $\tilde{S}$ and shuffle it. 
    \WHILE {the training stop condition is not met}
    \STATE Take a batch $\mathbf{s}_1$ from the set $S$.
    \STATE Take another batch $\mathbf{s}_2$ from the dataset $\tilde{S}$.
    \STATE Randomly sample $SNR_1$, $\gamma$, $\alpha$  individually.
    \STATE Set $SNR_2=SNR_1+\gamma$.
    \STATE Semantic encode and fuse $\mathbf{s}_1$ and $\mathbf{s}_2$ as $\mathbf{y}$ (tramsmitter).
    \STATE Transmit $\mathbf{y}$ over the two channels and users get $\mathbf{y}_1$, $\mathbf{y}_2$.
    \STATE Semantic de-fuse and decode $\mathbf{y}_1$ as $\mathbf{z}_1$ (worse user).
    \STATE Compute loss $L_1$ (worse user).
    \STATE Semantic de-fuse and decode $\mathbf{y}_2$ as $\tilde{\mathbf{z}}_1$ and $\mathbf{z}_2$ (better user).
    \STATE Compute $L_2$ (better user).
    \IF {$\alpha==0$}
    \STATE Set $\beta=0$ and then compute loss $L=L_2$.
    \ELSIF {$\alpha==1$}
    \STATE Compute loss $L=L_1$.
    \ELSE
    \STATE Compute loss $L=L_1+\lambda L_2$.
    \ENDIF
    \STATE Update all the parameters to minimize $L$.
    \ENDWHILE
	\end{algorithmic}
\end{algorithm}

\section{EXPERIMENTS}
In this section, we evaluate the performance of the proposed semantic communication scheme for DBC to transmit the image by using CIFAR-10 dataset. We use Adam optimizer to train the system for 100 epochs with a learning rate of $1\times10^{-4}$ and then train for another 50 epochs with a learning rate $1\times10^{-5}$. The batch size is 128. For the scalarization parameters, we set $\lambda=6$ and $\lambda\beta=0.1$. Without loss of generality, we use $l=2k$, $h=n=32$ and the bandwidth ratio $\frac{k}{h\times n\times 3}=0.25$ in the experiment. 
\begin{figure}
    \begin{center}
    \subfigure[Raw image $\mathbf{s}_1$]{
    \includegraphics[width=3.5cm]{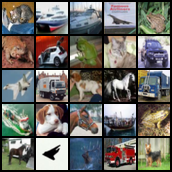}}
    \subfigure[Proposed]{
    \includegraphics[width=3.5cm]{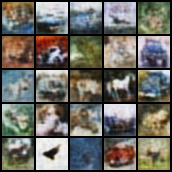}}
    \subfigure[TD]{
    \includegraphics[width=3.5cm]{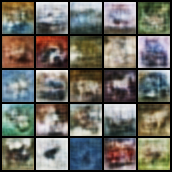}}
    \subfigure[PA]{
    \includegraphics[width=3.5cm]{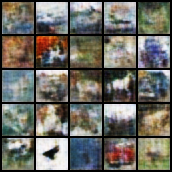}}] 
    \subfigure[Raw image $\mathbf{s}_2$]{
    \includegraphics[width=3.5cm]{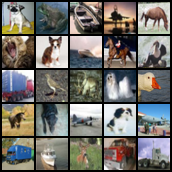}}
    \subfigure[Proposed]{
    \includegraphics[width=3.5cm]{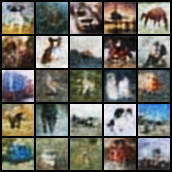}}
    \subfigure[TD]{
    \includegraphics[width=3.5cm]{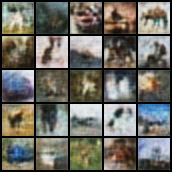}}
    \subfigure[PA]{
    \includegraphics[width=3.5cm]{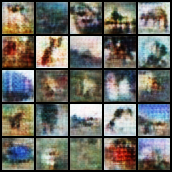}}
    \caption{Raw iamges and images reconstructed by different methods with CIFAR10 dataset.}
    \label{pic}
    \end{center}
\end{figure}

In the experiment, we consider the power allocation (PA) scheme and the TD scheme as benchmarks. For both benchmarks, $\mathbf{s}_i$ can pass the semantic encoder and then be fed into its own MLP, yielding the transmitted symbols for one user. For the PA scheme, the superposition coding with successive interference cancellation  is then exploited to produce the transmission symbols $\mathbf{y}$. For the TD scheme, the transmission symbols for different users are transmitted over their assigned time slots.

Fig. \ref{pic} depicts the visible results of the reconstructed images based on different methods. We use $\alpha =0.5$, $SNR_1=5~dB$ and $SNR_2=10~dB$. The upper image is for the worse user, while the lower image is for the better user. It can be observed that the proposed scheme produces the most clear recovered images among those based on the TD and PA schemes.

  \begin{figure}[t]
    \begin{center}
      \includegraphics*[width=8cm]{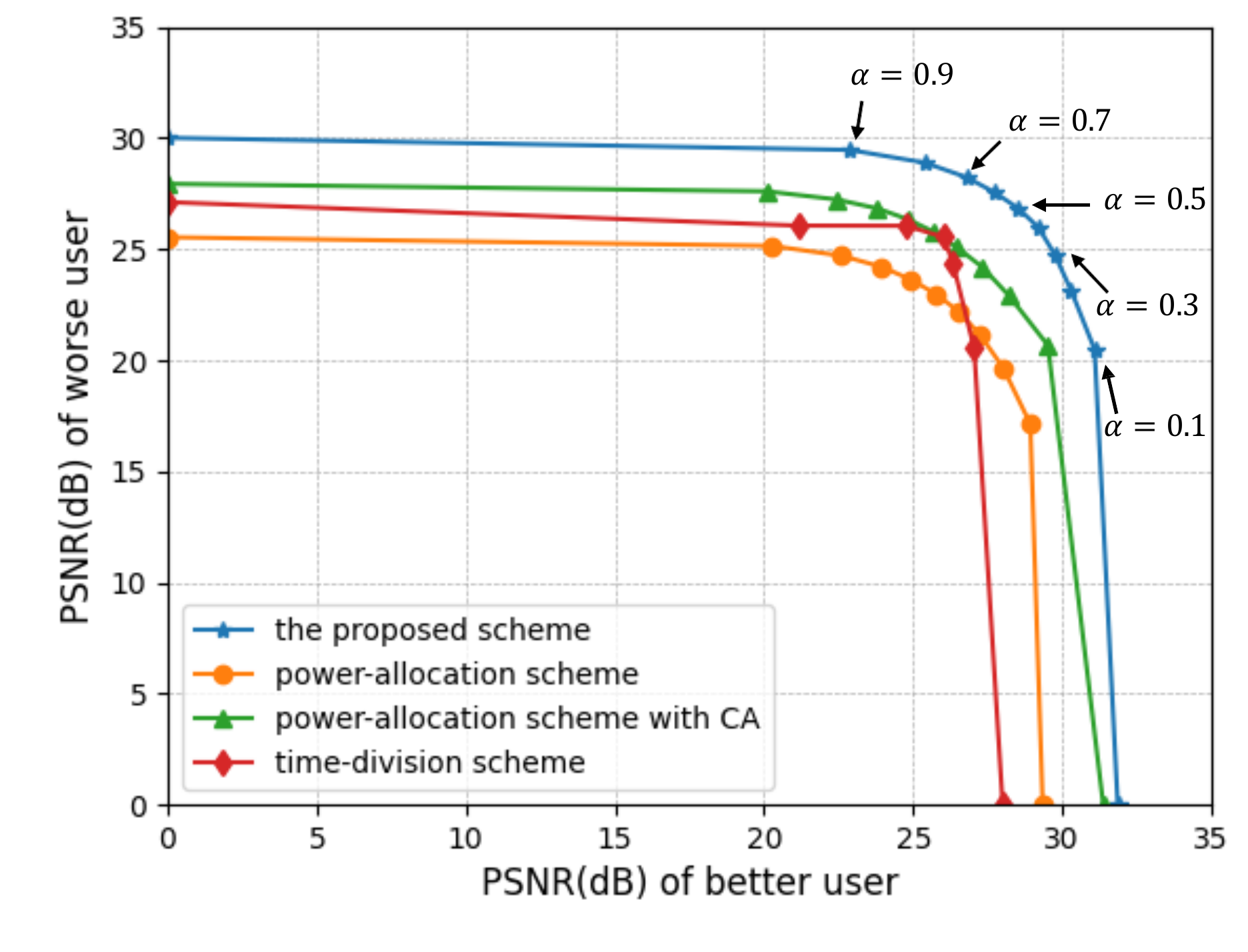}
    \end{center}
      \caption{\small{Semantic performance region for DBC with different methods. Here, the SNR of the worse/better user is $-5~dB$/$0~dB$.}}
      \label{SPR}
  \end{figure}

  \begin{figure}[t]
    \begin{center}
      \includegraphics*[width=8cm]{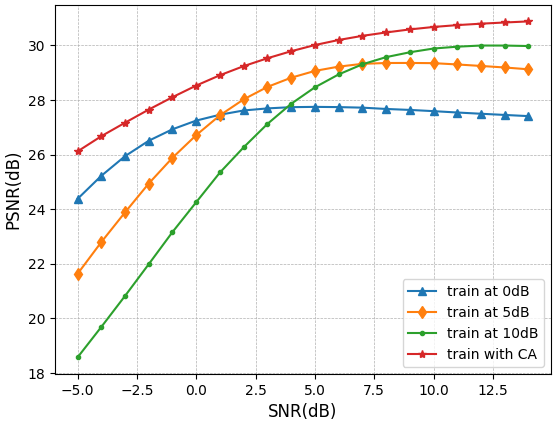}
    \end{center}
    \caption{\small{PSNR vs. SNR of the better user with the proposed schme. Here, $\alpha$ is 0.5.}}
    \label{DBCA_better}
  \end{figure}

  \begin{figure}[t]
    \begin{center}
        \includegraphics*[width=8cm]{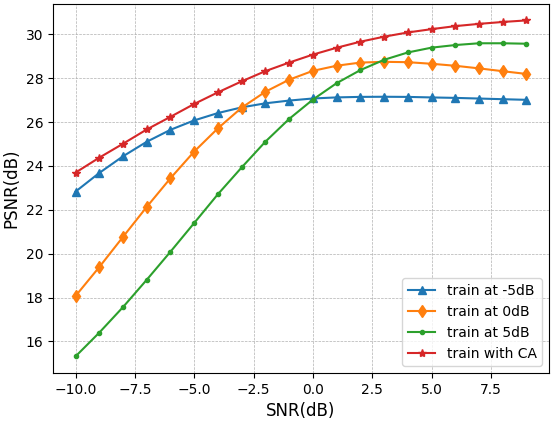}
    \end{center}
      \caption{\small{PSNR vs. SNR of the worse user with the proposed schme. Here, $\alpha$ is 0.5.}}
      \label{DBCA_worse}
  \end{figure}


Next, we evaluate the performance of DBC by using PSNR. The PSNR of a single user cannot reflect the comprehensive performance of DBC. Therefore, we can describe the achievable PSNR groups of both users, which form a region called the \emph{semantic performance region}. Fig. \ref{SPR} shows the semantic performance region with different schemes. For the proposed scheme, we can obtain different PSNR groups by adjusting fusion ratio $\alpha$, as shown in Fig. \ref{SPR}. It is seen that with the proposed scheme, the PSNR of the worse user increases with $\alpha$, but leads to the decrease of PSNR of the better user. This result matches (\ref{mlp_y_t}). We also clearly see that the semantic performance region of the proposed scheme strictly contains the region of other bechmarks. This fact shows that the proposed scheme can achieve the best performance for both users in DBC compared to traditional methods. The PA scheme without channel adaptive (CA) has the smallest region, which indicates that the power allocation scheme is not suitable in the semantic communications system. The gap between the power allocation scheme without CA and that of with CA reveals the CA gain. The gap between the power allocation scheme with CA and the proposed scheme reveals the fusion gain.

Fig. \ref{DBCA_better} and Fig. \ref{DBCA_worse} show PSNR vs. SNR for the better user and the worse user based on the proposed scheme, respectively.We can observe that the proposed FISF module incorporating CSI into semantic features to adapt the channel can provide significant performance gain. For example, for the better user at $SNR=-5$ dB the system training at -5 dB achieve the best performance than that of other SNR-fixed training, e.g. 0 dB. However, it is still 1.29 dB lower in the PSNR performance than proposed scheme. 

\section{CONCLUSION}
In this paper, we have proposed a novel semantic communications system for wireless image transmission over two-user degraded broadcast channels. The transmitter can extract the semantic features of two users' images and fuse these semantic features into a joint latent representation for broadcasting. We have designed a flexible image semantic fusion scheme that dynamically controls the weight of semantic features in the joint latent representation and adapts the respective semantic features to the respective channels with different SNRs. Experimental results have shown that the proposed system significantky dominates the traditional methods, such as TD and PA, for wireless image transmission over two-user degraded broadcast channels.
 
\footnotesize
\bibliographystyle{IEEEtran}
\bibliography{reference}{}
\end{document}